\documentclass[conference]{IEEEtran}
\IEEEoverridecommandlockouts
\usepackage{cite}
\usepackage{amsmath,amssymb,amsfonts}
\usepackage{graphicx}
\usepackage{textcomp}
\usepackage{xcolor}
\usepackage{soul}
\usepackage{algorithm}
\usepackage{algpseudocode}
\usepackage{makecell}
\usepackage{diagbox}
\usepackage{pifont}
\usepackage{hyperref}
\usepackage{soul}
\usepackage{subcaption}
\def\BibTeX{{\rm B\kern-.05em{\sc i\kern-.025em b}\kern-.08em
    T\kern-.1667em\lower.7ex\hbox{E}\kern-.125emX}}

\newcommand{\commentout}[1]{\iffalse #1 \fi}

\setstcolor{brown}

\begin{document}


\title{Drop-Connect as a Fault-Tolerance Approach for RRAM-based Deep Neural Network Accelerators\\}

\author{\IEEEauthorblockN{
Mingyuan Xiang\IEEEauthorrefmark{2}\IEEEauthorrefmark{1}\thanks{\IEEEauthorrefmark{1} These two authors contribute equally.}, Xuhan Xie\IEEEauthorrefmark{2}\IEEEauthorrefmark{1},
Pedro Savarese\IEEEauthorrefmark{3}, Xin Yuan\IEEEauthorrefmark{2},
Michael Maire\IEEEauthorrefmark{2} and Yanjing Li\IEEEauthorrefmark{2}}
\IEEEauthorblockA{
\IEEEauthorrefmark{2}Department of Computer Science,
University of Chicago, Chicago, IL USA, 60637\\
}
\IEEEauthorblockA{
\IEEEauthorrefmark{3}Toyota Technological Institute at Chicago, Chicago, IL USA, 60637\\
}
}

\maketitle

\begin{abstract}
Resistive random-access memory (RRAM) is widely recognized as a promising emerging hardware platform for deep neural networks (DNNs). 
Yet, due to manufacturing limitations, current RRAM devices are highly susceptible to hardware defects, which poses a significant challenge to their practical applicability.
In this paper, we present a machine learning technique that enables the deployment of defect-prone RRAM accelerators for DNN applications, without necessitating modifying the hardware, retraining of the neural network, or implementing additional detection circuitry/logic.
The key idea involves incorporating a drop-connect inspired approach during the training phase of a DNN, where random subsets of weights are selected to emulate fault effects (e.g., set to zero to mimic stuck-at-1 faults), thereby equipping the DNN with the ability to learn and adapt to RRAM defects with the corresponding fault rates.
Our results demonstrate the viability of the drop-connect approach, coupled with various algorithm and system-level design and trade-off considerations. We show that, even in the presence of high defect rates (e.g., up to 30\%), the degradation of DNN accuracy can be as low as less than 1\% compared to that of the fault-free version, while incurring minimal system-level runtime/energy costs.

\end{abstract}

\begin{IEEEkeywords}
  fault tolerance, neural network, machine learning, RRAM
\end{IEEEkeywords}

\section{Introduction} \label{sec:intro}

\begin{figure}[htbp]
  \centerline{\includegraphics[width=\columnwidth]{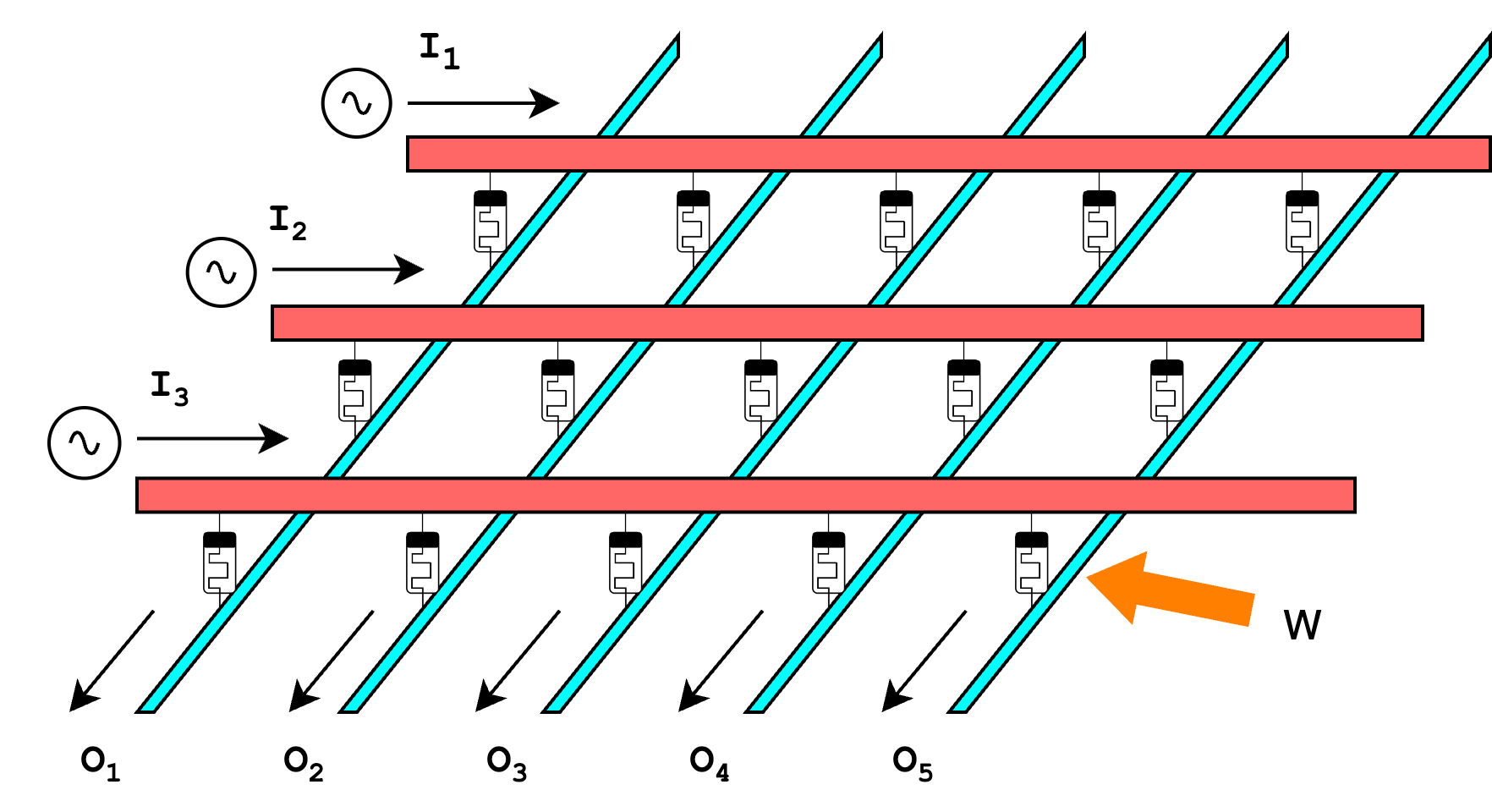}}
  \caption{Example of a RRAM crossbar.}
  \label{crossbar}
\end{figure}

Recent advancements in Deep Neural Networks (DNNs) have demonstrated significant success across various applications. However, the increasing complexity and capabilities of DNNs necessitate substantial computational power and memory bandwidth in conventional Von Neumann architectures to accelerate DNN applications. 
A promising alternative lies in the utilization of novel architectures constructed with emerging technologies. Among the various options, the Resistive RAM (RRAM) crossbar-based architecture, comprised of memristor cells \cite{isaac}, emerges as an innovative compute-in-memory solution that not only reduces power consumption, but also boosts processing speeds. Illustrated in Fig. \ref{crossbar} is a standard RRAM crossbar. Within the crossbar, DNN kernels are unfolded and embedded with each memristor cells, each of which retaining a single weight value, while input data is continuously streamed into the crossbar from its wordlines.
The analog nature of this architecture makes it well-suited for vector-matrix multiplication, as the dot product operation can be replicated using Kirchhoff’s circuit law.

While RRAM-based DNN accelerators offer dramatic improvements in energy efficiency and throughput over traditional architectures, memristor devices are prone to various hardware defects \cite{defect}, primarily due to manufacturing constraints. The most prominent among these defects manifest as the Stuck-At-Faults (SAF). In particular, the Stuck-At-One (SA1) faults, where memristor cells are trapped in a high state, can lead to both write and read failures so that weight values stored in these faulty cells are always interpreted as 0's. SA1 faults account for a significant portion (9.04\%) of all faults in RRAM crossbars \cite{defect}. 



Even if only a fraction of these faults occur, the impact on DNN model accuracy can be significant. Existing work \cite{retrain, map, chenchen} typically relies on the actual defect distributions that can be obtained through memory testing \cite{defect, marchc} to design fault-tolerant RRAM accelerators. However, these techniques either compromise network accuracy (e.g., as seen in remap techniques \cite{map}, especially with higher defect rates) or impose substantial deployment overhead (e.g., in retrain-based approaches, where the entire model must be trained from scratch every time it is deployed to a new RRAM crossbar). Moreover, because of the inherent variability and drift associated with the forming and switching operations in memristors, the distribution of defects in a crossbar can evolve over time, which adds another layer of complexity and makes it even more challenging and less attractive to deploy previously proposed solutions. Checksum-based techniques \cite{checksum} inspired by algorithm-based fault-tolerance are effective and efficient for RRAM DNN fault tolerance as well. However, these techniques require additional hardware resources to compute and compare checksums, which are not always feasible or desirable.

This paper explores a training method inspired by a machine learning training technique called drop-connect. In drop-connect \cite{dc}, randomly selected sets of weights of a DNN are set to 0 (and not updated) during different training iterations. While this technique was originally proposed to avoid over-fitting during training, it can also be used to emulate random SA1 faults for a given fault rate. This approach is a natural solution to enable RRAM crossbars to tolerate SA1 faults. The intuition is that, if the network can be trained to recognize and understand various fault effects and patterns, taking into account the corresponding fault distributions, then it should be well-equipped to handle similar fault effects and distributions during the inference phase. 

The most notable benefits of our approach is that it does not require modifications to the hardware, retraining of the neural network, or the implementation of additional detection circuitry/logic, such as checksum logic. Moreover, this approach offers the opportunity to train the network so that it can adapt to the evolving defect distributions overtime, further enhancing its effectiveness and applicability.

The major contributions of our work are: 
\begin{itemize}
\item We conduct a thorough investigation and analysis of the drop-connect inspired technique as a fault-tolerance solution for RRAM-based DNN accelerators. More importantly, our technique requires \textbf{no information} of actual defect distribution at deployment. The key differences between our work and others are summarized in Table.~\ref{tab:diff}
\item We provide the detailed implementation of this approach.
\item We perform experiments across various representative DNNs and fault rates, and explore various tradeoffs between network accuracy and efficiency (in terms of system-level runtime and energy consumption) to demonstrate the efficacy of our approach.  
\end{itemize}

\begin{table}[htbp]
  \caption{Comparison of major fault-tolerant techniques for RRAM crossbars.}
  \label{tab:diff}
  \centering
  \begin{tabular}{|c|c|c|}
  \hline
  method & Additional Circuits & Retraining and Mapping \\ 
  \hline
  Liu et al.~\cite{chenchen} & Defect Detection & required \\ 
  \hline
  Das et al.~\cite{checksum} & Checksums & None \\
  \hline
  Li et al.~\cite{conf/iccd/LiWLL19} & Refresh and Detection & required \\ 
  \hline
  Ours & \textbf{None} & \textbf{None} \\
  \hline
  \end{tabular}
\end{table}

Our key findings and results are summarized below.
\begin{enumerate}
\item The drop-connect-inspired approach is a viable solution for enabling fault-tolerant RRAM DNN accelerators, particularly when the fault rate is relatively low or when a modest accuracy loss (e.g., $5\%$) is acceptable. For certain networks (e.g., MobileNet V2), the degradation in accuracy is less than 1\% even when the fault rate is very high (30\%), while incurring no additional system-level costs.

\item  The best accuracy levels are achieved when a higher drop-connect rate is used, compared to the expected RRAM fault rate. However, high drop-connect rates (beyond $30\%-40\%$) adversely affect network accuracy even in fault-free scenarios.

\item  Further improving accuracy in RRAM crossbars with drop-connect-inspired fault tolerance may involve widening the original network (increasing the number of channels). This compensates for information loss due to weights forced to certain values, but at the cost of higher runtime and energy consumption, and diminishing returns. Systematically exploring trade-offs between system-level costs and network accuracy is therefore crucial. Our results show that, a 20\%/60\% increase in the number of channels yields up to 4\%/12.5\% improvements in test accuracy, respectively, compared to 0\% increase, while incurring up to 42.6\%/153.3\% performance/energy costs.

\item Due to the unique properties of convolution layers with 1x1 kernels, running these layers in traditional architectures such as CPUs achieve allows the network to achieve higher network and runtime/energy efficiency simultaneously.

\item In certain networks, modifying the structure of a few critical layers serves as an alternative approach to accompany drop-connect in order to improve DNN accuracy. For example, in ResNet20 \cite{resnet}, increasing the kernel size in the shortcut layers from 1x1 to 3x3 and deploying these layers in RRAM crossbars achieves comparable network accuracy to deploying the original network layers in fault-free devices. 

\end{enumerate}

Our approach is orthogonal to other fault-tolerance techniques such as the post-training remapping strategies, and they can be combined to further enhance DNN accuracy. Moreover, our work has brought to light an important revelation -- there exist nuances in the adaptation of machine learning approaches to tackle system-level challenges, which necessitates an in-depth understanding not only of the original machine learning technique, but also of the broader system-level implications and tradeoffs associated with applying an existing method for a new purpose. For example, based on key result \ref{dropconnectrate} discussed above, the ``sweet-spot'' for the drop-connect rates must be carefully selected to obtain the optimal accuracy, while employing the vanilla drop-connect technique where the drop-connect rate is set to SA1 fault rate is likely to be sub-optimal.
This revelation holds broader relevance beyond our specific study, extending to other system/design challenges where machine learning techniques are adopted and adapted as solutions, for which there is an imperative need for meticulous consideration and thorough analysis. 

The rest of the paper is structured as follows: Sec.~\ref{sec:metho} describes our methodology. Sec.~\ref{sec:res} presents our experimental results, followed by related work in Sec.~\ref{sec:related}  and conclusions in Sec.~\ref{sec:con}.

\section{Methodology}~\label{sec:metho}

\subsection{The Original Drop-Connect Approach}

Drop-connect \cite{dc} is a regularization technique to prevent DNN models from overfitting during training.
A convolution operation integrated with drop-connect can be defined as:
\begin{equation}
    O_{mij}^{(l)} = \frac{1}{1-p} \sum_{n} \sum_{k} \sum_{s} I_{n, i+k, j+s}^{(l)} \cdot W_{mnks}^{(l)} \cdot M_{mnks}^{(l)}
\end{equation}
Here, $I_{nij}^{(l)}$, $O_{mij}^{(l)}$, and $W_{mnks}^{(l)}$ are inputs, outputs, and weights of the convolution layer $l$. $M^{(l)}$ is a mask that satisfies the Bernoulli distribution, i.e., $M^{(l)} \sim \text{Bernoulli}(p)$. $p$ is the drop-connect probability. Additionally, the output needs to upscale by $\frac{1}{1-p}$ during training to maintain the expected distribution.

For networks that incorporate normalization layers, such as batch normalization \cite{bn}, there is an important implementation detail that is worth noting. A scaling factor of $\frac{1}{1-p}$ must also be applied to normalize the weights when drop-connect is applied \cite{dc}. This ensures that only non-zero weights are normalized to preserve the correct weight value distribution. 

\subsection{Drop-Connect Adapted for Fault-Tolerance Purposes}~\label{sec:dropconnect}
As discussed in Sec.~\ref{sec:intro}, drop-connect provides a way for DNN models to learn to compensate for RRAM defects regardless of the actual defect cell distribution while maintaining high accuracy. As such, no additional costs and overheads are required during deployment time.

An important question for adapting drop-connect for fault-tolerance purposes is the following: what is the optimal drop-connect rate during training? To answer this question, we sweep the drop connect rate for a given fault rate (see Sec.~\ref{sec:res}). In these experiments, as the running statistics (i.e., moving mean and variance) of batch normalization can differ between the drop-connect rate and the actual defect rate, our training algorithm involves an additional epoch to align the scaling factor of normalization layers with the SA1 fault rate during inference (which is different from the drop-connect rate during training). As shown in Algorithm~\ref{algo:dc_last_epoch}, during this epoch, we apply the inference-time scaling factor (i.e., RRAM SA1 fault rate) to adjust the moving mean/variance values, matching them more closely to the actual SA1 fault distribution, while keeping the network weights constant. 

\begin{algorithm} 
\caption{UpdateVar(MODEL)}
\label{algo:dc_last_epoch}
\begin{algorithmic}[1]
    \State // $p'$: inference-time scaling factor, i.e., SA1 fault rate
    \For{$p'$ in [0\%, 10\%, 20\%, 30\%]}
        \State {MODEL.train()}
        \For {\textit{batch} in TrainingSet}
            \State {Freeze MODEL.weights}
            \State {MODEL.dropConnect($p'$)}
            \State {MODEL.forward(\textit{batch})}
        \EndFor
    \EndFor
\end{algorithmic}
\end{algorithm}

\begin{figure}[htbp]
  \centerline{\includegraphics[width=\columnwidth]{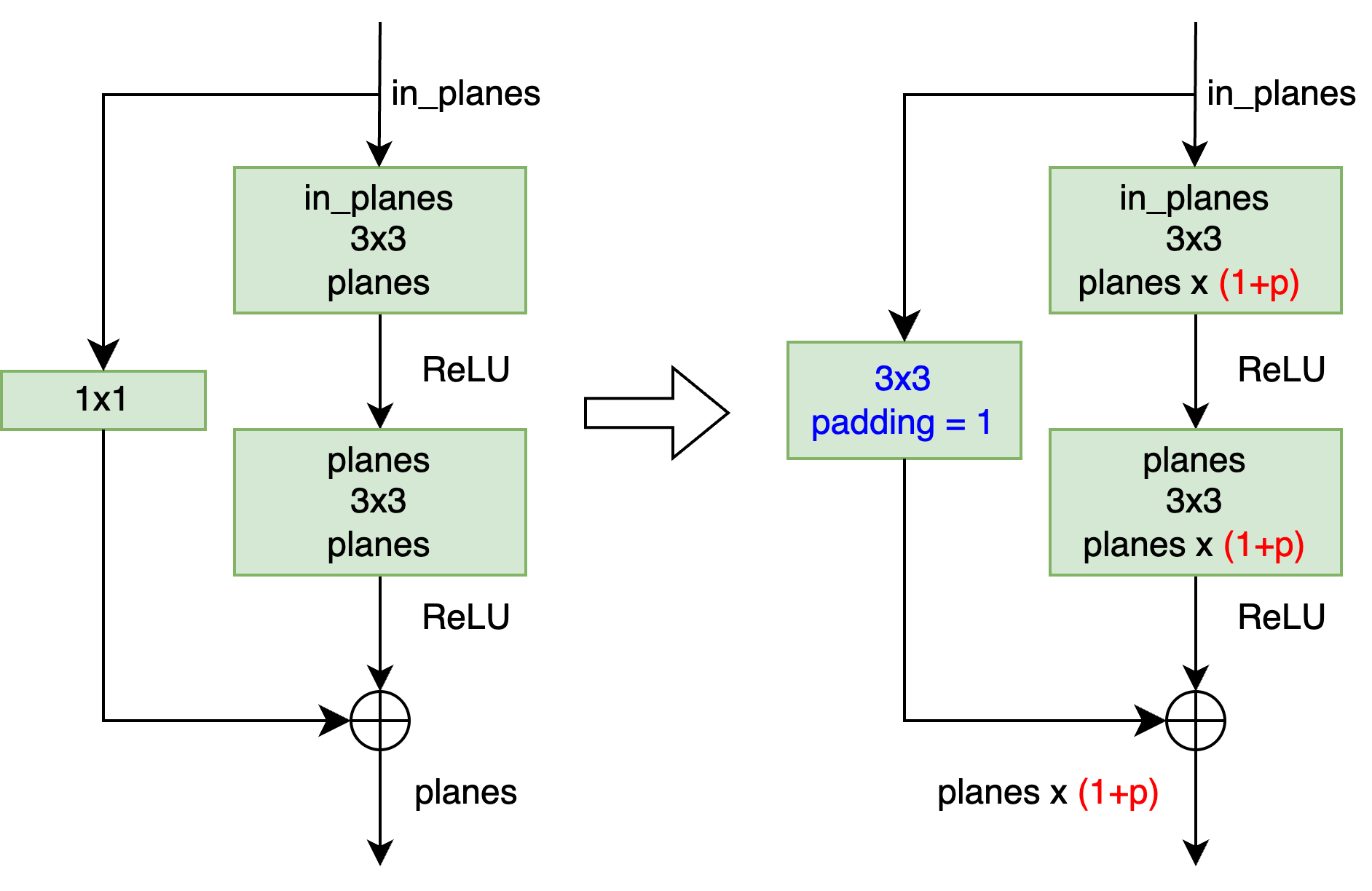}}
  \caption{Increasing network width (red) and/or increasing kernel size of 1x1 convolution layers (blue) to compensate for information loss due to drop-connect. }
  \label{inc_ch}
\end{figure}

Moreover, we explore widening the original network (i.e., increasing the number of channels in convolution layers) to compensate for information loss due to drop-connect, as shown in Fig.~\ref{inc_ch} (highlighted in red). $p$ is the percentage increase in the number of channels.

\subsection{Special Considerations on Convolution Layers with 1x1 Kernels} \label{1x1conv}
When applying the drop-connect approach during training, it is not applied to convolution layers where the kernel size is 1x1. Such 1x1 convolution layers are prevalent in modern DNN models, such as point-wise convolution in MobileNet V2 \cite{mobilenet}, or shortcut convolution in ResNet20 \cite{resnet}. Unlike convolutions with kernel sizes larger than 1x1, these 1x1 convolution layers can be efficiently optimized on traditional processors, such as CPUs. Specifically, 1x1 convolutions are equivalent to multiplication and accumulation (MAC) operations along the input channel dimensions. These MAC operations can be readily computed using optimized architecture extensions such as instructions in the Intel Advanced Vector Extensions (AVX) \cite{avx}.
Furthermore, loop unrolling can be employed to mitigate the loop branching overhead. Consequently, 1x1 convolution layers can be executed efficiently on traditional processors, so it is not critical to offload them to accelerators such as RRAM crossbars.  

Coincidentally, these 1x1 convolution layers also constitute a critical component of the network model. For instance, the point-wise convolution in MobileNet V2 increases the output channels for the separable convolution. When an SA1 fault occurs, the corresponding output channels are all zeroed, leading to a significant loss of information. At the same time, drop-connect does not work well in these layers, again due to the significant loss of information (results in Sec.~\ref{sec:res}).

In summary, for networks with 1x1 convolution layers, considering both system-level runtime/energy efficiency and network accuracy, it is more advantageous to execute these 1x1 convolution layers on traditional fault-free devices, such as CPUs and GPUs. 

\subsection{Modifying Network Layer Structure to Enhance Drop-Connect} \label{3x3shortcut}
Motivated by the promising application of the drop-connect approach for fault-tolerance and the crucial role observed in 1x1 convolution layers for neural network robustness, we explore an alternative machine learning technique which involves modifying the structure of the 1x1 convolution layers. The idea is to increase the kernel size -- for example, from 1x1 to 3x3. In the original network, the drop-connect-inspired approach encounters challenges with the 1x1 convolution layers, primarily stemming from notable information loss within these layers. By expanding the kernel size, we can now apply drop-connect to these layers, enabling their deployment on RRAM-based DNN accelerators. Fig~\ref{inc_ch} provides an example of how the 1x1 shortcut layer is modified (highlight in blue). In this example, each side of the input is padded with 0, and the 1x1 convolution is substituted with a 3x3 convolution. Note that, these 3x3 shortcut layers serve the same role as identity mapping in residual neural networks.


\section{Evaluation and Key Results} \label{sec:res}

\subsection{Experiment Setup}~\label{cell}

We demonstrate the effectiveness of our approach through simulation experiments. We simulate the RRAM crossbar behaviors and fault effects using an in-house PyTorch-based \cite{pytorch} simulator. In our simulation, each memristor cell stores an 8-bit weight value, a common RRAM crossbar configuration ~\cite{yao2020fully, DBLP:conf/glvlsi/WangTXLGYL015, DBLP:conf/dac/HeLEYF19}.

We included a set of representative DNN benchmarks in our experiments, including ResNet20 \cite{resnet}, MobileNet V2 \cite{mobilenet} and VGG13 \cite{VGG} on the CIFAR-10 dataset \cite{cifar}. Given a fault rate, after training these networks using the drop-connect-based approach discussed in Sec.~\ref{sec:metho}, we construct 100 RRAM crossbars with random faults following the given fault rate, and apply the trained model on each of the crossbars. We collect and report the test accuracy by taking the average of these 100 runs.

We performed a systematic analysis by sweeping fault rates and drop-connect rates, increasing the number of channels, and experimenting with expanding the kernel size of short-cut layers in ResNet20 from 1x1 to 3x3.

\subsection{Results for Adapting Drop-Connect as a Fault-Tolerance Solution} \label{dropconnectrate}
\begin{figure}[h!]
    \centering
    \includegraphics[width=0.48\textwidth]{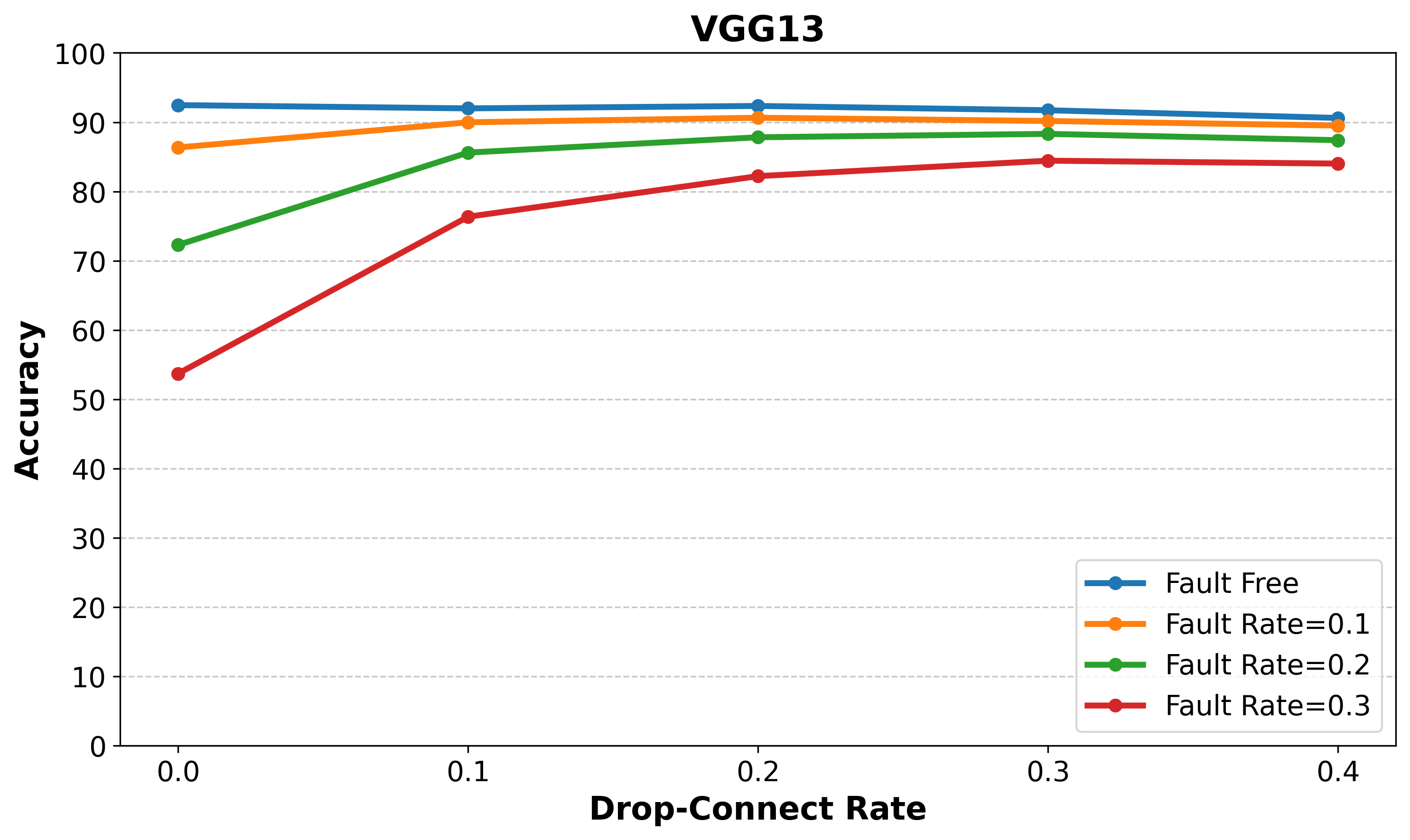}
    \caption{Network Accuracy of VGG13 for different drop-connect and fault rates.}
    \label{fig:VGG13}
    
    \vspace{0.3cm}
    
    \centering
    \includegraphics[width=0.48\textwidth]{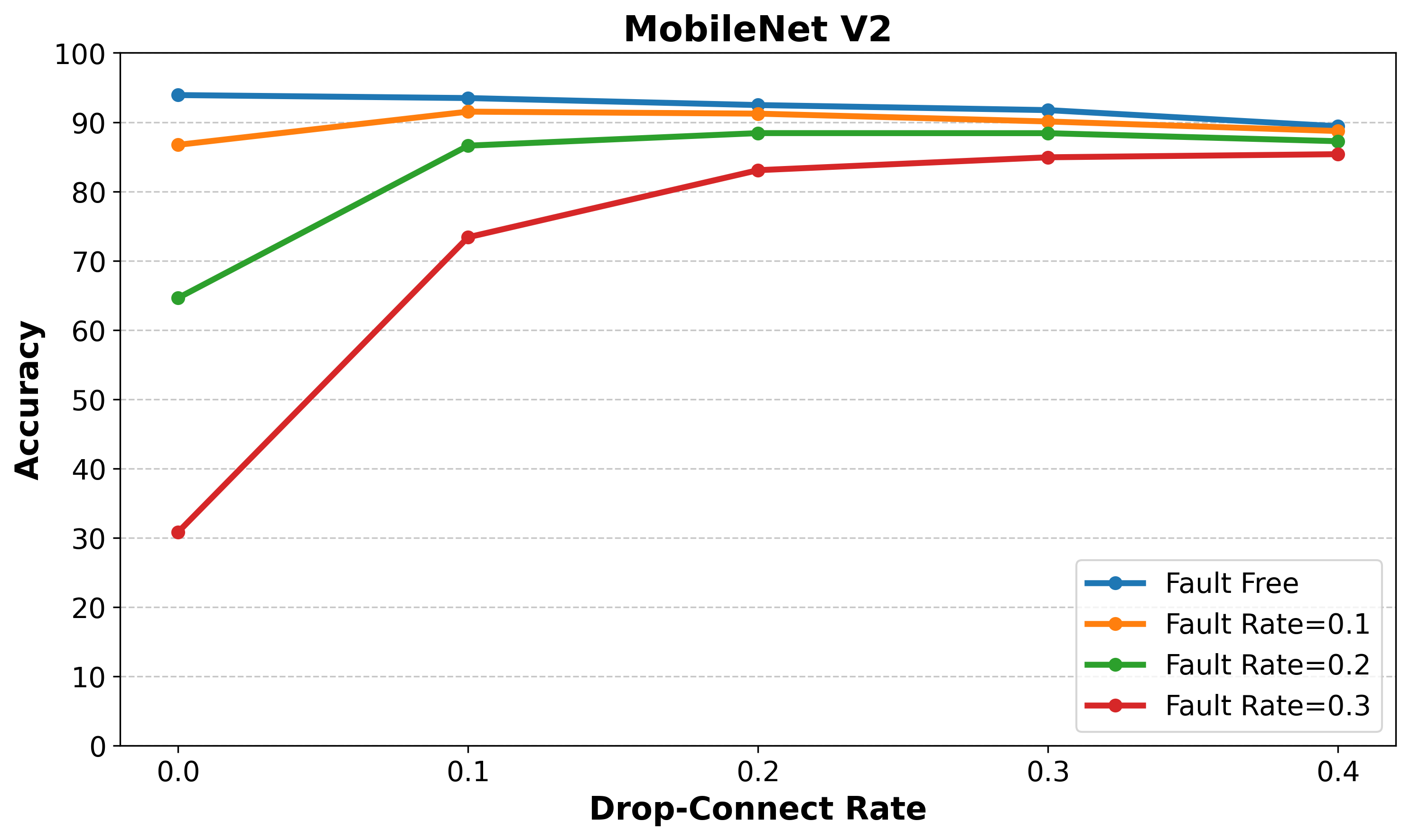}
    \caption{Network Accuracy of MobileNet V2 for different drop-connect and fault rates.}
    \label{fig:MobileNetV2}

    \vspace{0.3cm}

    \centering
    \includegraphics[width=0.48\textwidth]{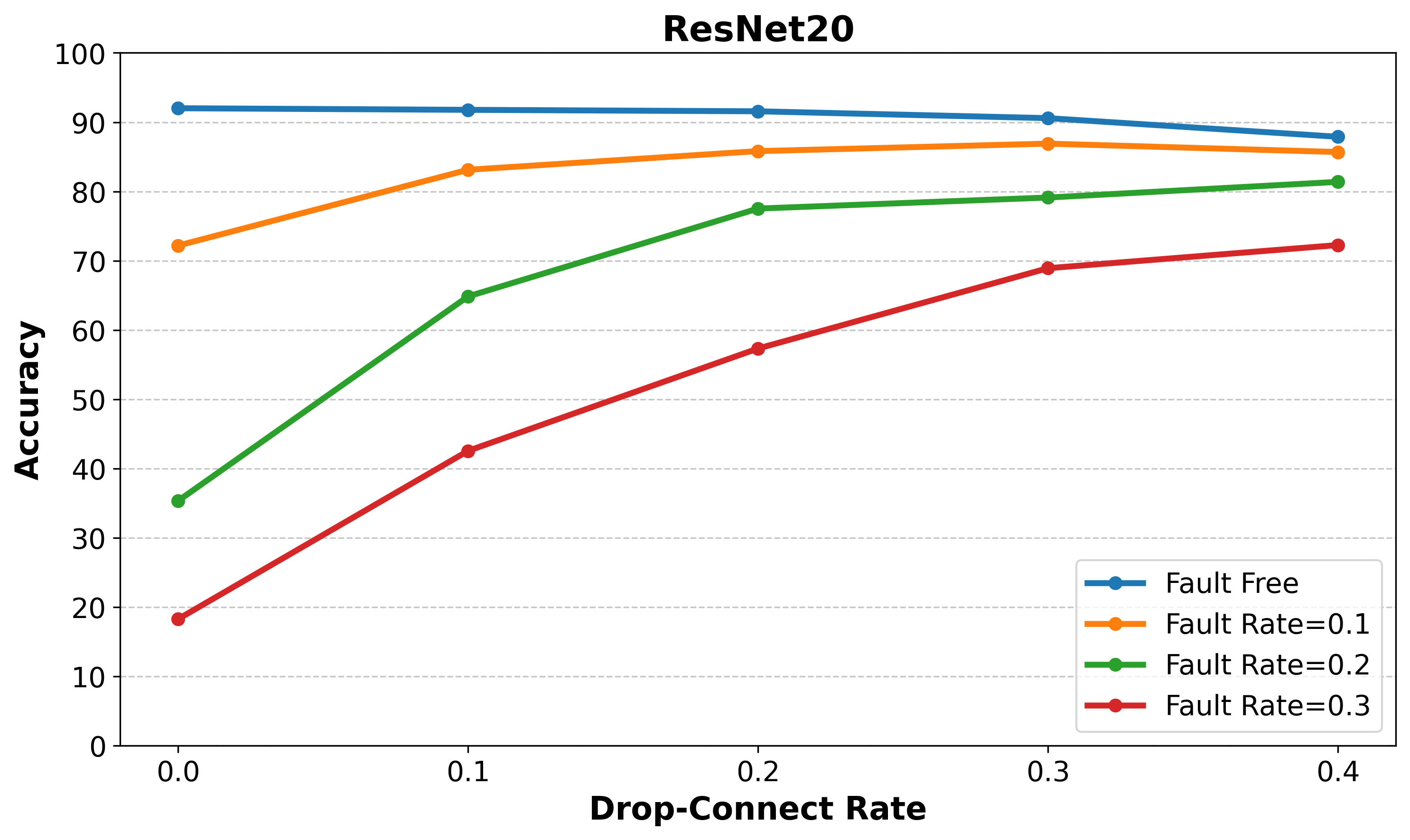}
    \caption{Network Accuracy of ResNet20 for different drop-connect and fault rates} 
    \label{fig:ResNet20}
\end{figure}

The accuracy results for different drop-connect and fault rate combinations are shown in Figs. \ref{fig:VGG13}, \ref{fig:MobileNetV2}, and \ref{fig:ResNet20} for different networks.

Two key observations can be made from these results. First, the drop-connect-based approach is effective in enabling fault-tolerant RRAM-based DNN accelerators, when the SA1 fault rate is  up to 10\%. For example, for a 10\% fault rate,  the degradation in network accuracy is less than 2\% for VGG and MobileNet V2. For higher fault rates, larger gaps in accuracy are seen -- 4-10\%/6-20\%  for a 20\%/30\% fault rate, respectively. However, this approach still provides substantial benefit in model accuracy compared to networks training without drop-connect (i.e., the data points correspond to 0\% drop-connect rate).

\begin{figure*}[ht!]
    \centering
    \begin{subfigure}[b]{0.32\textwidth}
        \centering
        \includegraphics[width=\textwidth]{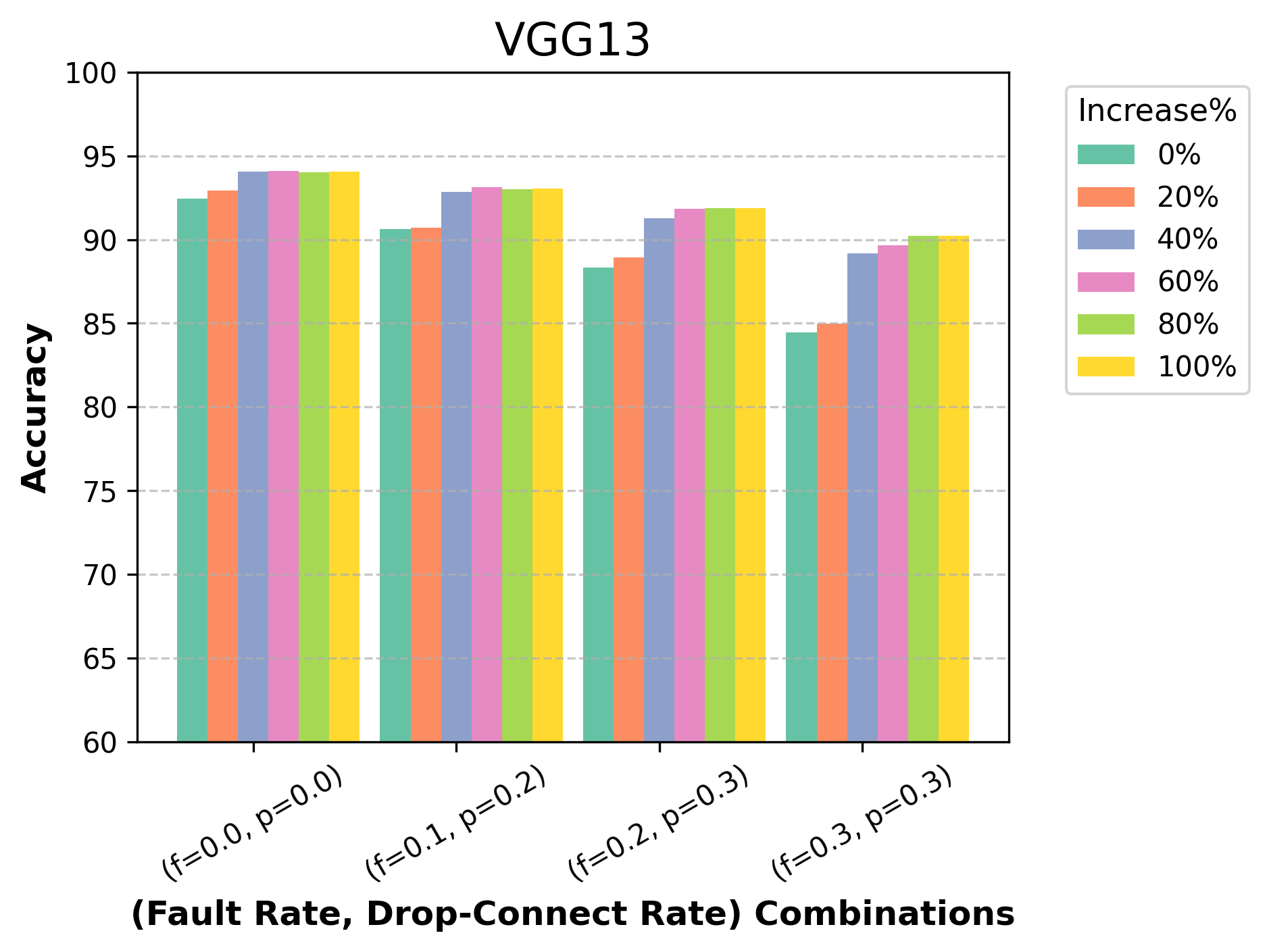}
        \caption{VGG13}
        \label{fig:VGG13_bar}
    \end{subfigure}
    \begin{subfigure}[b]{0.32\textwidth}
        \centering
        \includegraphics[width=\textwidth]{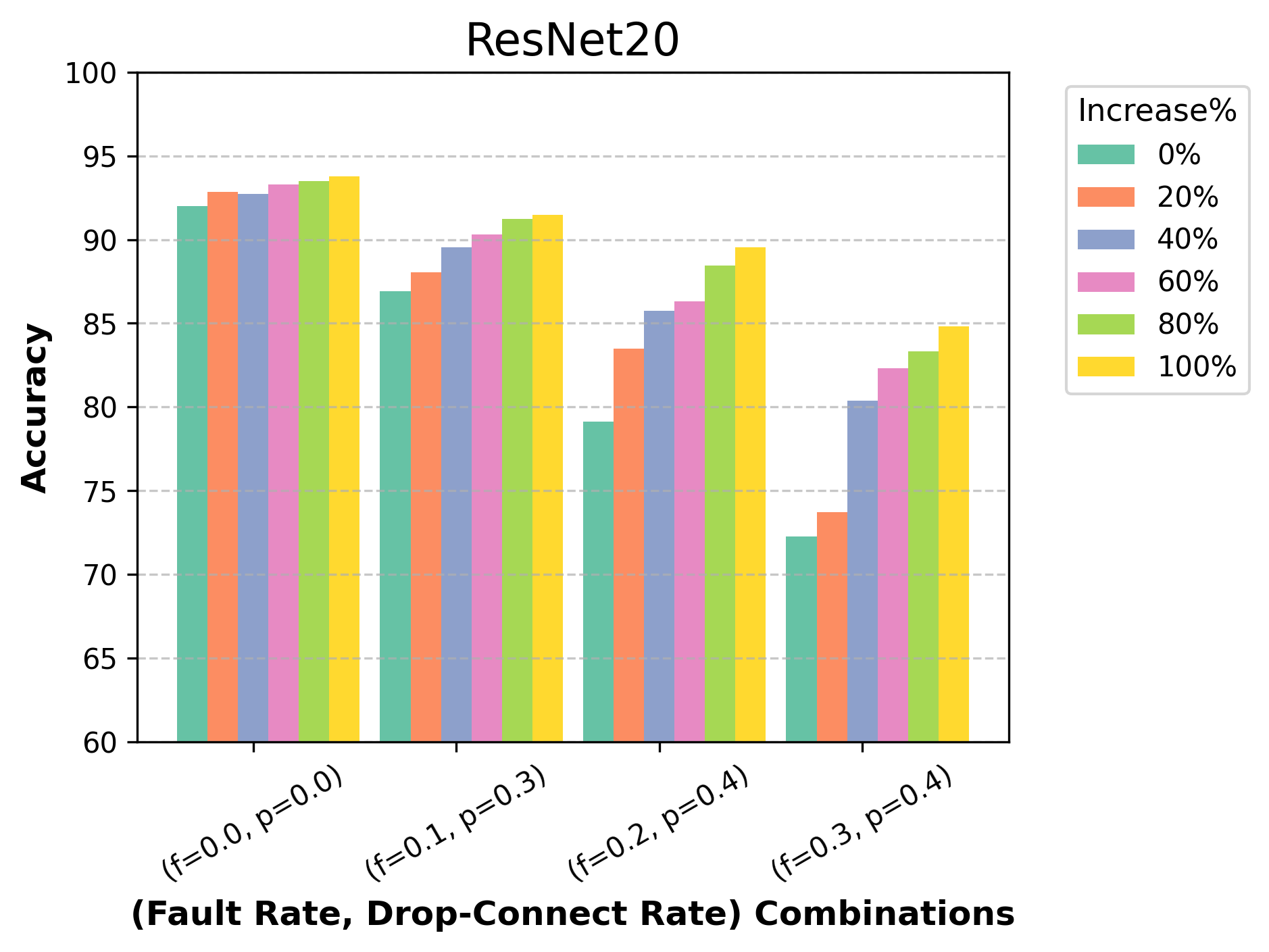}
        \caption{ResNet20}
        \label{fig:ResNet20_bar}
    \end{subfigure}
    \begin{subfigure}[b]{0.32\textwidth}
        \centering
        \includegraphics[width=\textwidth]{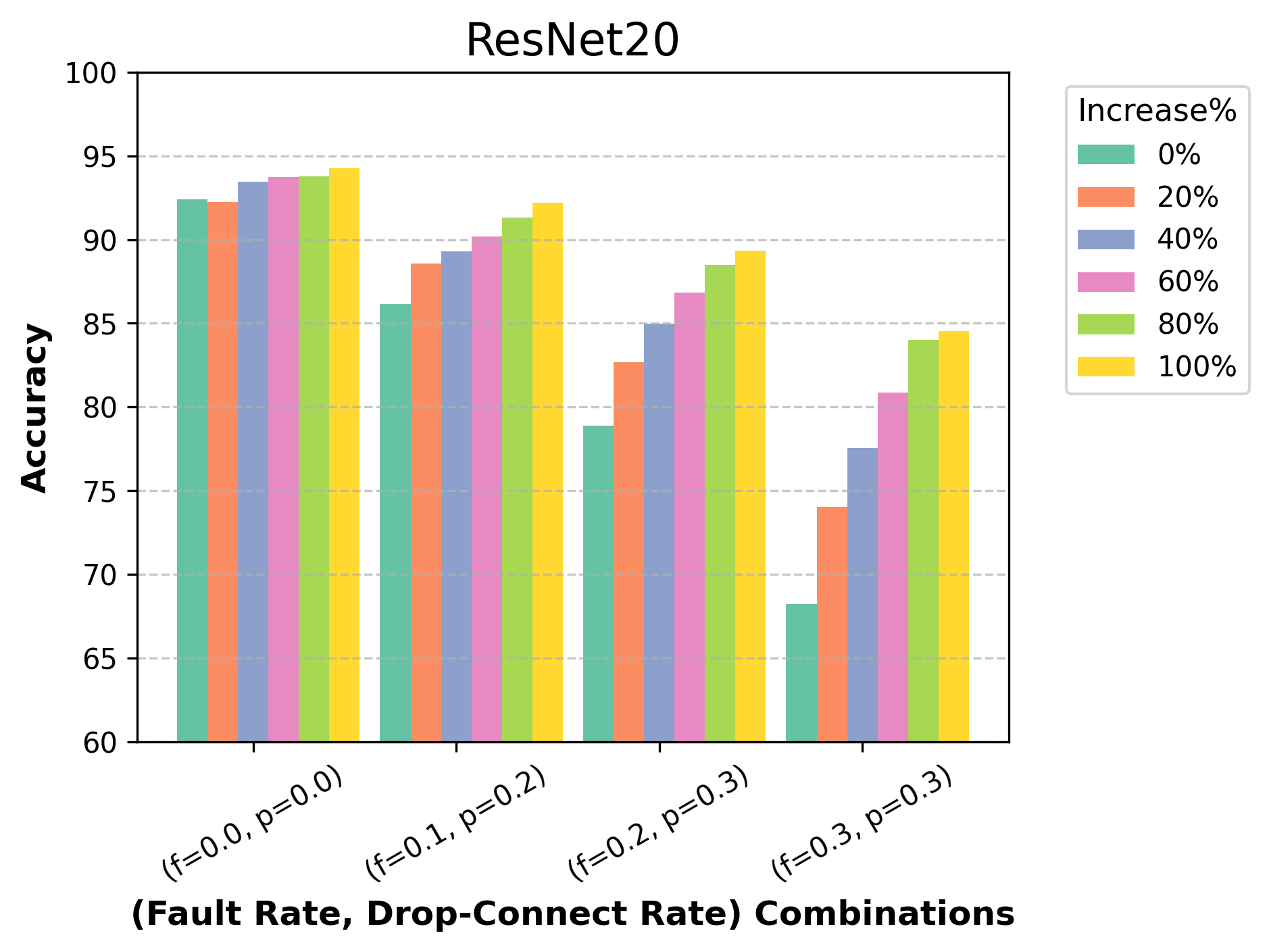}
        \caption{ResNet20 (3x3 short-cut kernels)}
        \label{fig:ResNet20_3x3}
    \end{subfigure}
    \caption{(a) and (b) the combinations of the fault rate and the drop-connect rate that achieve the highest network accuracy for different network width; (c) increasing the size of kernels in short-cut layers from 1x1 to 3x3.}
\end{figure*}

Second, the best accuracy for each fault-rate is achieved when the drop-connect rate is higher. The best drop-connect rate for a given fault-rate is shown in Figs. \ref{fig:VGG13_bar} and \ref{fig:ResNet20_bar}. 

Overall, our approach works well for VGG, since it is an over-parameterized network with inherently higher fault-tolerance than other networks. For MobileNet V2, our approach achieves similar levels of effectiveness. This may be attributed to executing the critical point-wise layers in fault-free hardware. On the other hand, for ResNet20
which is a more challenging network, the drop-connect approach alone may not be able to achieve the desired level of accuracy.  


Network accuracy improves as we increase the width of the network, as shown in Figs. \ref{fig:VGG13_bar} and \ref{fig:ResNet20_bar}. Note that, we did not run the same experiments for MobileNet V2 because the accuracy is already very high even with the original network configuration. 

More channels come with more filters in convolution layers, allowing the model to learn a greater number of features from the data. In addition, having more channels provides the model with more flexibility to determine which filter is crucial for the task, which leads to higher accuracy.
This method is particularly pronounced for ResNet20. Its impact becomes more evident when evaluating the model on RRAM crossbars with higher fault rates. With a 10\%/20\% fault rate, the degradation in accuracy compared to the fault-free ResNet20 is only 0.2\%/3\%, respectively.  
VGG also benefits from this method, achieving at least 90\% accuracy ($<2\%$ degradation) on RRAM crossbars with fault rates ranging from 0 to 30\%. 

However, there is a trade-off here. Increasing the number of channels inevitably leads to higher computational costs and longer runtime. We estimate the energy consumption and latency with respect to the width of networks, using the computational efficiency and power efficiency results from the 64-chip ISAAC RRAM accelerator \cite{isaac}, and the results are shown in Table.~\ref{tab:perf}. The width of each network is normalized to that of the original model (first row in the table).

\begin{table}[htbp]
  \caption{Latency and energy estimation for different network configurations with different network widths (normalized to the original network width).}
  \label{tab:perf}
  \resizebox{\columnwidth}{!}{
  \scalebox{0.5}{
  \begin{tabular}{|c|c|c|c|}
    \hline
     & ResNet20 & MobileNet V2 & VGG13 \\
    \hline
    Width of Network & \multicolumn{3}{c|}{latency ($s$), energy ($kW$)} \\
    \hline
     1x & \makecell{15.74 \\ 113.32} & \makecell{0.68 \\ 4.89} & \makecell{87.83 \\ 632.16} \\
    \hline
     1.2x & \makecell{22.13 \\ 159.28} & \makecell{0.96 \\ 6.93} & \makecell{125.25 \\ 901.56} \\
    \hline
     1.4x & \makecell{29.80 \\ 214.53} & \diagbox{}{} & \makecell{170.59 \\ 1227.84} \\
    \hline
     1.6x & \makecell{39.11 \\ 281.56} & \diagbox{}{} & \makecell{222.53 \\ 1601.71} \\
    \hline
  \end{tabular}
  }}
\end{table}

Increasing network width can also lead to potential overfitting problems as well as diminishing returns. For instance, we observe that the benefit of increasing network width for VGG starts to diminish at 40\% increase in the number of channels, and increasing the network width further does not significantly improve accuracy. Moreover, for ResNet20, with a fault rate of 30\%, the accuracy degradation of around 7\% may still be too high, even though it is a substantial improvement over the network accuracy achieved using the original network configuration (the accuracy degradation is around 79\%).  

Therefore, given the various tradeoffs associated with this approach, it is crucial to systematically explore and select the best design point that balances network accuracy and runtime/energy costs for a given use scenario. It is also worth noting that, even with higher system-level costs, our approach can be still more desirable than others that require additional hardware support (e.g., special RRAM circuit or peripheral checksum logic), which requires more design efforts and cannot be readily deployed on existing hardware.

\subsection{Results for Increasing the Size of Kernels}
In Fig.~\ref{fig:ResNet20_3x3}, we show the accuracy for ResNet20 when the kernel size of the short-cut layers is increased from 1x1 to 3x3, with drop-connect applied to these layers during training. The results are comparable to those shown in Fig.~\ref{fig:ResNet20_bar}, suggesting that this is a promising alternative approach to achieve fault tolerance.

\subsection{Results Demonstrating the Fault Criticality of Convolution Layers with 1x1 Kernels}
\begin{figure}[h!]
    \centering
    \includegraphics[width=0.48\textwidth]{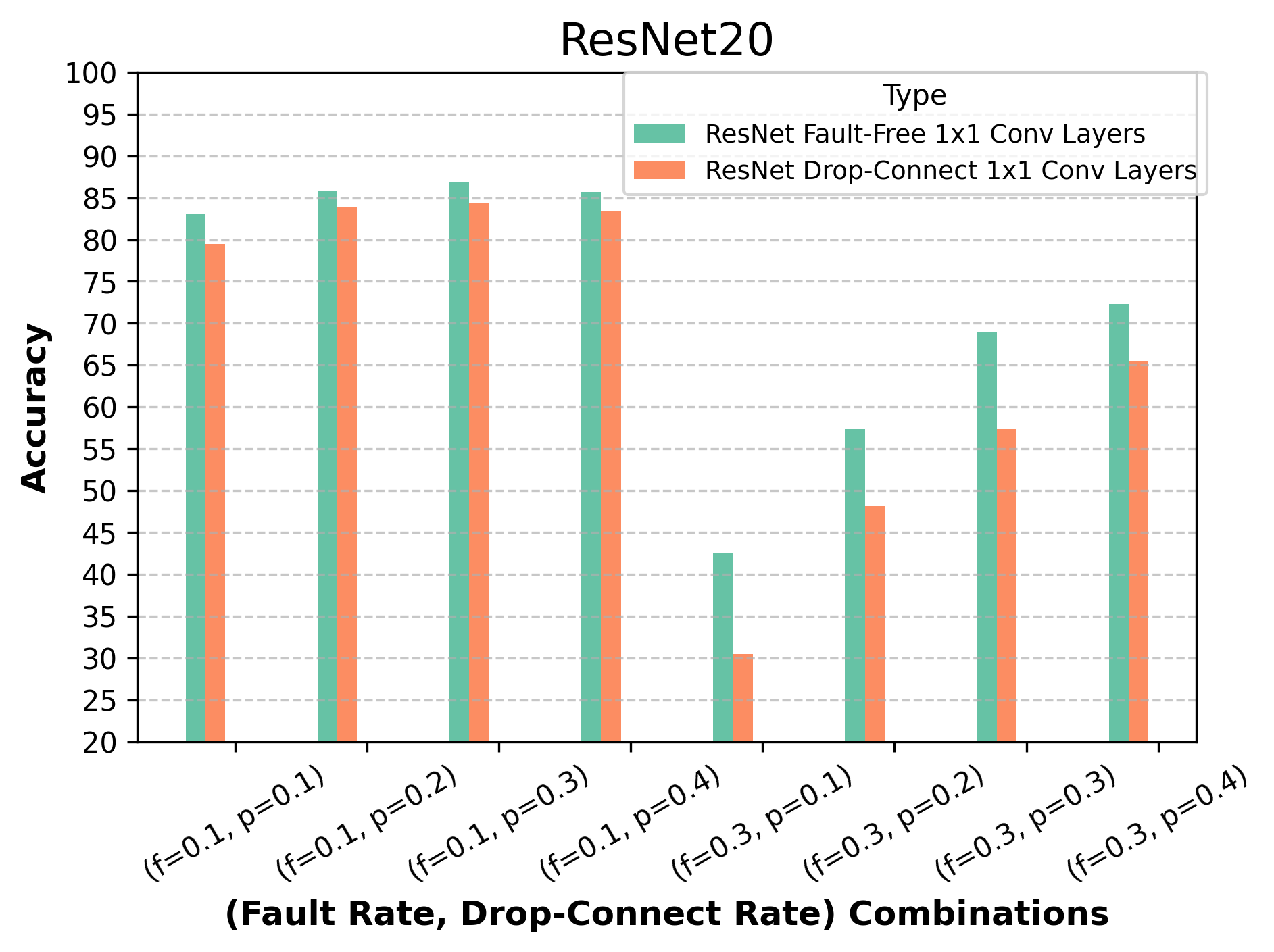}
    \caption{ResNet20, comparison of fault-free 1x1 convolution layers and when drop-connect and SA1 faults are applied to the same layers.}
    \label{fig:ResNet_Comparison}
    \vspace{0.5cm}
\end{figure}

In Fig.~\ref{fig:ResNet_Comparison}, we demonstrate the fault criticality of convolution layers with 1x1 kernels by comparing the difference between: (1) our approach, which is to execute these layers in traditional fault-free architectures, and (2) applying drop-connect and mapping these layers to faulty RRAM crossbars. The dramatic difference is clear evidence that executing these layers in a fault-free manner is crucial to achieve high network accuracy.

\section{Related Work} \label{sec:related}

RRAM defects and faults have been extensively examined and characterized in previous work \cite{defect, softerrors, sneakpath}.
They can be divided into two main categories: soft errors \cite{softerrors} and hard errors \cite{defect}.
In the scope of this paper, the focus is on hard errors, because solutions at the circuit level have already been
devised to address soft RRAM errors \cite{softerrors}.
Hard errors include the stuck-at and transition faults. Although traditional methods
such as the March-C algorithm \cite{marchc} or the squeeze-search algorithm \cite{defect} are effective in detecting hard RRAM errors, they cannot mitigate the defects \cite{DBLP:conf/itc/ChaudhuriC18}. Thus, fault-tolerance techniques are required for practical deployment of RRAM accelerators. 

A retrain and post-mapping method is proposed by~\cite{chenchen}, where they detect the defect distribution
and retrain the DNN models. After training, a remapping scheme is adopted such that the least important
weights are mapped to faulty memristor cells. Similar retrain techniques are also proposed in \cite{retrain, map, conf/dac/XiaLNCW17, conf/iccd/LiWLL19}. A significant limitation of retrain-based techniques is that they require training an entire neural network from scratch every time the network is deployed to a new accelerator (or when the fault distribution of an accelerator changes), which incurs prohibitively high resource overheads, and may not even be possible because the training dataset may not be available. Post-mapping methods avoid the high overhead of retrain, but they alone cannot achieve acceptable levels of network accuracy. 

Multiple circuit-level solutions have been proposed to handle RRAM defects. For example, \cite{checksum} employs a parity matrix for a majority-vote-based checksum across the entire crossbar. And in \cite{ecc}, an approach based on the diagonal error correction code is proposed. However, such circuit solutions are only effective when the defect rate is no higher than 5\% \cite{DBLP:conf/itc/LiuXWC18}. For higher defect rates, they introduce non-negligible latency overheads (26\% \cite{ecc}) or memory overheads ranging from 9\% to 30\% \cite{checksum}.

The concept of drop-connect was mentioned in \cite{dropins} to achieve fault-tolerance in RRAM crossbars;
however, the authors approached its application naively, lacking the profound understanding
and comprehensive system-level trade-off analysis that our work provides — a necessity for the successful
adaptation of this technique. Consequently, their work lacks meaningful insights and falls short in delivering
desirable results. Notably, results are only reported for a limited set of simple and shallow models
(e.g., LeNet \cite{lenet} and AlexNet \cite{alexnet}), yet acceptable levels of network accuracy still cannot be achieved -- the training accuracy of MNIST is 78.34\%, a significant degradation vs. the fault-free case where a 92.8\% training accuracy can be achieved. A pruning-based scheme was proposed by~\cite{conf/isqed/YuanLMCKSFLZPLR21} under
the intuition that defects can be mitigated if the positions of pruned weights
overlapped with faulty memristor cells. However, this method also suffers from significant network accuracy degradation with high fault rates. For example, the accuracy of ResNet18 using the CIFAR-10 dataset drops to less than 20\% even for a low fault rate of 10\% in~\cite{conf/isqed/YuanLMCKSFLZPLR21}.

Drop-connect has been proposed to mitigate the effects of hardware defects on DNNs in systolic array-based DNN accelerators~\cite{journals/tcad/OzenO22}. 
Moreover, a dropout-based method, where neurons are dropped (instead of weights being dropped in the case of drop-connect) is proposed to overcome faults in the neurons of spiking neural networks ~\cite{conf/date/SpyrouEACLS21}. In contrast, our work focuses on RRAM-based DNN accelerators. 


\section{Conclusion} \label{sec:con}
In this paper, we perform a thorough study on a drop-connect-inspired technique to enable fault tolerance in RRAM-based DNN accelerators. Distinct from previous work, we incorporate various algorithm-/system-level considerations and analysis, and also conduct comprehensive experiments. Through our study, we have obtained various new insights, and the main conclusion is that our approach is viable as a fault-tolerance solution, especially if the fault rate is low and/or if modest network accuracy degradation is acceptable. Our approach allows various tradeoffs between network accuracy and system-level runtime/energy efficiency to be obtained, so that the best design point can be chosen for the specific use scenarios. At the same time, this approach does not require modifications to the hardware, retraining of the neural network, or the implementation of additional detection circuitry/logic. However, in order to tolerate a higher fault rate or close the accuracy gap, other machine learning and system techniques are needed. We plan to build on top of this work and investigate even more efficient fault-tolerance techniques targeting RRAM-based DNN accelerators.  


\bibliographystyle{IEEEtran}

\bibliography{references}
\vspace{12pt}

\end{document}